Physical Sciences: Chemistry

**Ultra-High Yield Growth of Vertical Single-Walled Carbon Nanotubes:**

**Hidden Roles of Hydrogen and Oxygen**

Guangyu Zhang,[1] David Mann,[1] Li Zhang,[1] Ali Javey,[1] Yiming Li,[1] Erhan Yenilmez,[1] Qian Wang,[1] James McVittie,[2] Yoshio Nishi,[2] James Gibbons[2] and Hongjie Dai[1]*

[1]Department of Chemistry and Laboratory for Advanced Materials, Stanford University, Stanford, CA 94305.

[2]Department of Electrical Engineering, Stanford University, Stanford, CA 94305

* Correspondence to hdai@stanford.edu

Tel: 650-723-4518

Fax: 650-725-0259

**Manuscript Information:** 14 pages of text and 6 figures

Abstract: 104 words

Manuscript: 24,038 characters

**Abstract**

An oxygen assisted hydrocarbon chemical vapor deposition (CVD) method is developed to afford large-scale highly reproducible ultra high-yield growth of vertical single-walled carbon nanotubes (SWNT). It is revealed that reactive hydrogen (H)-species, inevitable in hydrocarbon-based growth, are damaging to the formation of $sp^2$-like SWNTs. The addition of oxygen scavenges H-species and provides a powerful control over the C/H ratio to favor SWNT growth. The revelation of the roles played by hydrogen and oxygen leads to a unified and universal optimum growth condition for SWNTs. Further, a versatile method is developed to form vertical SWNT films on any substrate, lifting a major substrate-type limitation for aligned SWNTs.

**Introduction**

Carbon nanotubes in aligned forms(1-6) are interesting for various scientific and practical applications ranging from electronics to biological devices. While several CVD methods have been developed to grow vertically aligned multi-walled carbon nanotubes (V-MWNT), growth of vertical single-walled carbon nanotubes (V-SWNTs) is still in an early stage.(7, 8) The formation of nanotube vertical films is indicative of high yield growth from densely packed catalytic particles. That is, almost every particle produces a nanotube and bundling of neighboring tubes leads to collective vertical growth. A key element in a recent ethylene CVD method that has obtained V-SWNT growth is the addition of ~150 ppm water vapor.(8) Another method that has obtained V-SWNT growth is an alcohol CVD method using ethanol as the carbon feedstock. In both cases, the key role played by the oxygen-containing species is suggested to be oxidation of amorphous carbon to facilitate SWNT growth.

Though we have made an effort to utilize both the water-assisted and alcohol CVD growth methods, we have been unable to obtain vertically aligned SWNTs, which indicates the narrow parameter space and growth window of these methods. Also, the specific roles played by the oxygen-containing species remain speculative and require further investigation. A more detailed understanding could lead to improved and more reproducible growth conditions. Here, we present a molecular oxygen assisted plasma-enhanced CVD (PECVD) growth of V-SWNTs at the full 4-inch wafer scale. We find that adding oxygen (~1%) to methane in the PECVD affords highly reproducible growth of densely packed SWNTs. Control experiments and knowledge from previous diamond PECVD work suggest that the key role played by oxygen in the high yield SWNT growth is to balance C and H radicals, and specifically to provide a C-rich and H-deficient condition to favor the formation of $sp^2$-like graphitic SWNT structures. We reveal that reactive hydrogen species are generally unfavorable to SWNT formation and growth and can etch pre-formed SWNTs. The negative role of hydrogen and the positive role of oxygen have important implications to SWNT growth in general by various methods. Lastly, we present a simple but powerful method to form V-SWNT films for the first time on any desirable substrate (including metals and plastics) with strong interfacial adhesion.

**Methods**

Our nanotube synthesis was carried out in a conventional 4-inch thermal CVD system with an inductively coupled radio-frequency (RF, 13.56 MHz) plasma source located near the entrance of the growth gases.[9] The substrates used were $SiO_2/Si$ with nominally 1 to 2Å-thick Fe films (by quartz crystal thickness monitoring) deposited by electron beam evaporation. The thin Fe film was first annealed in oxygen at 550°C for 10 min and then heated in hydrogen to the growth temperature of 720°C. This treatment produced nearly a monolayer of Fe clusters on $SiO_2$ with an average diameter of ~1.3 nm as estimated from atomic force microscopy (AFM) measurements. We found that the formation of dense and relatively uniform particles was essential for V-SWNT growth. During nanotube growth, the compositions of gases in the tube-furnace were methane (~66%), hydrogen (~12%), oxygen (~ 1%) and Ar (~21% as carrier gas) with a total pressure of 0.3-0.4 Torr. The gas flow rates were $CH_4/H_2/O_2$=160sccm/30sccm/2.4sccm (standard cubic centimeter per minute). Ar was used as carrier gas for $CH_4$ and $O_2$. The percentage of partial pressures of various gases followed $CH_4$: $H_2$: $O_2$ =66% : 12% : 1% (The rest is Ar). The RF plasma was generated at a power of 60-70 W for 10 to 30 min for nanotube growth. This condition was highly reproducible in growing vertical SWNTs from run to run and day to day.

**Results and Discussions**

Our method produced nearly a monolayer of Fe clusters on 4-inch wafer scale $SiO_2$ (Fig. 1a) substrates with an average diameter of ~1.3 nm as estimated from atomic force microscopy (AFM) measurements. We found that the formation of dense and relatively uniform particles was essential for V-SWNT growth. With the sub-monolayer catalyst seeds and ~1% oxygen added to our CVD, uniformly black nanotube films were grown on full 4-inch wafers (Fig. 1 b). Scanning electron microscopy (SEM) revealed that the films consisted of nanotube strands oriented vertically to the substrate (Fig.1c&1d). The length of the nanotubes is ~ 10 μm for 10 min growth and scaled approximately with growth time. Transmission electron microscopy (TEM) revealed exclusively SWNTs without any MWNTs or double-wall tubes (Fig. 1f). The TEM grids were prepared by sonicating a V-SWNT covered substrate in a solvent and then drop-drying the suspension on the grid. Raman spectroscopy (Fig. 1e) of the as-grown samples identified resonant radial breathing modes (RBM) of nanotubes in the range of 132cm$^{-1}$ to 230cm$^{-1}$, corresponding to nanotubes 1 to 2 nm in diameter.(10) The clear separation between the G peaks at ~1580cm$^{-1}$ and ~1610cm$^{-1}$ (Fig. 1e) is also characteristic of SWNTs.(10) By forming densely packed catalytic seed particles in lithographically defined regions shaped in squares, circles or strips, we grew V-SWNTs to form square towers, circular towers or sheets replicating the shape of the catalytic regions (Fig.2) with the thickness of the sheets down of ~ 100nm (Fig. 2a).

Our V-SWNT synthesis results are highly reproducible and the optimum $O_2$ concentration of ~1% can be easily controlled experimentally. As mentioned earlier, the vertical orientation of SWNTs is indicative of high yield growth with close to one-to-one growth of nanotubes from the seed particles. Through control experiments (Fig. 3a, 3b) with varying concentrations of gases at a fixed temperature and total pressure, several growth trends emerged and shed light on the role played by oxygen in high yield nanotube formation. Without the addition of $O_2$, the same dense-catalyst monolayer failed to produce high yield of SWNTs packing into vertical films under any conditions tested. Without the $H_2$ flow on the other hand, vertical SWNTs can still be grown using $CH_4/O_2$ (Fig. 3a). Growth in $CH_4/H_2$ gave extremely low yield of SWNTs (Fig. 3b). These results indicated that H-rich conditions do not favor SWNT growth.

We suggest that an important role played by oxygen in enhancing SWNT growth is the removal of reactive H radicals (relative to C-species) that exist in hydrocarbon based growth of nanotubes, and that high concentrations of H species do not favor the formation and growth of $sp^2$-like SWNTs[11] (Fig.3). We first note that the $CH_4/H_2/O_2$ gases have been used previously for $sp^3$ diamond synthesis, albeit under $H_2$ rich (>90%) and low $CH_4$ (several percent) conditions.[11, 12] Optical emission spectroscopy (OES, Fig.3c inset) has established that adding oxygen to the $CH_4/H_2$ plasma removes the highly reactive H radicals via $H + O_2 \rightarrow OH + O$ with a large rate constant of $k_{i0} \sim 10^{17} cm^3/mol \cdot s$.[11] Oxygen can also remove and convert C-species into CO via various reactions, but with lower rate constants (~$10^{14} cm^3/mol \cdot s$).[11] Addition of $O_2$ to $CH_4/H_2$ CVD can therefore provide an effective route to tune the ratio between C· and H· species. It has been found that the C·/H· ratio needs to be below a limit (H·-rich) for the formation of $sp^3$ structure and high C·/H· ratios favor $sp^2$ carbon formation. The $sp^2$ and $sp^3$ carbon formation regimes can be controlled by changing oxygen concentration.(11) These and our SWNT growth results under varying $H_2$ and $O_2$ conditions indicate that high yield synthesis with $CH_4/H_2/O_2$ or $CH_4/O_2$ is due to oxygen removing H·, which removes or graeatly reduces the negative effect of H species on the growth (Fig.3c).

Our results lead to the conclusion that a favorable synthesis condition for SWNTs should employ sufficient C-feedstock with few or no reactive H-species. This may be a dilemma for hydrocarbon based synthesis of SWNTs since H· is an inevitable product of hydrocarbon decomposition. This dilemma can potentially be addressed with controlled addition of oxygen to scavenge H·, enhancing the C·/H· ratio and thus favoring $sp^2$ carbon production. High concentrations of reactive H· species are unfavorable to SWNT formation and growth likely due to attacking of the $sp^2$ C by H· to form $sp^3$ structures,(13, 14) giving low yield growth of SWNTs (Fig.3). Notably, hydrogenation and etching of SWNTs by H radicals generated in a plasma is known to occur even at room temperature.(13) To confirm hydrogen attacking of SWNTs, we carried out control experiments and observed etching of pre-formed SWNTs by hydrogen plasma (Fig. 4) under various conditions ranging from room temperature to our typical growth temperature. This result provided direct evidence of the negative effect of reactive H species to the structures of SWNTs.

It is tempting to suggest that oxidizers play a cleansing role by etching amorphous carbon and thus maintaining high catalyst activity for SWNT growth, as suggested by the water and alcohol assisted growth work.(7, 8) This is not the key effect, at least in our case, since we find that amorphous carbon deposition can be avoided by using suitable $CH_4$ concentrations via Ar dilution. Without oxygen, the yield of SWNTs (with any appreciable length) is low, suggesting the importance of oxygen in the initial nanotube nucleation and formation stage and not just during the sustained growth stage. Increased $H_2$ presence always leads to systematic decrease in SWNT yield, with or *without* oxygen presence (see Supplementary Information). In our case, due to plasma assisted decomposition, any $H_2$ leads to very high H· concentrations, much more so than in thermal CVD. Thus, the observed extremely low yield of SWNTs for $CH_4/H_2$ growth ($H_2<10\%$) strongly suggests the negative blocking effect of H-species to SWNT growth. With this negative H-effect identified, we conclude that the favorable enhancement effect of oxygen is via removal of H-species (Fig. 3a vs. Fig. 3b). Indeed, our optical emission spectroscopic measurement under our optimum $CH_4/H_2/O_2$ PECVD clearly identified significant OH species (Fig. 3c inset), lending direct spectroscopic evidence of the reaction of oxygen with H-species in our SWNT growth condition.

H-scavenging by oxygen species could also be a factor in the high yield SWNT growth by other methods(7, 8) although the precursors of oxidizing species differ. We carried out a control experiment to elucidate the effect of hydrogen to the growth yield of SWNT in the alcohol CVD growth process. We observed that increasing the $H_2$ concentration while keeping the alcohol vapor pressure constant systematically reduces the yield of nanotubes (Fig. 5). This provides evidence that hydrogen rich environments are also undesirable and have negative effects to the yield of SWNTs in standard thermal CVD. Further, a connection can be made with non-hydrocarbon based SWNT synthesis methods such as carbon monoxide CVD.(15, 16) CO-CVD without the involvement of hydrogen indeed produces high yield of SWNTs especially under high temperature and pressure when sufficient C-feedstock is obtained.(15) Other H-free high-yield growth of SWNTs includes laser ablation and arc-discharge that vaporizes solid carbon without involving any hydrogen.

A major difference between PECVD and thermal CVD is the much higher concentration of reactive radicals in the former. Small variations in the concentrations of molecules in PECVD can strongly affect the outcome of SWNT growth, as shown here (Fig. 3). For thermal CVD, the effect of varying the hydrogen concentration is much less pronounced than in the PECVD case but still noticeable. Two advantages of PECVD over thermal CVD are the efficient decomposition of gas molecules and the fact that the concentration of *reactive* species can be sensitively tuned by the precursor concentration. Previously, many groups have carried out PECVD synthesis of carbon nanostructures but have not succeeded in producing high yield SWNTs. We suggest that an origin of the growth failure is the unobvious negative role played by reactive H species abundant in hydrocarbon PECVD. We believe that by adding suitable amounts of oxygen to the various types of PECVD systems, and by using dense and relatively uniform catalyst particles, one should be able to readily produce SWNTs at high yields. Oxygen assisted PECVD could thus become a powerful and widely used method for efficient production of SWNTs. Note that we have attempted adding $O_2$ (0.04-4 percent partial pressure) to regular thermal CVD of hydrocarbons(17, 18) without achieving successful growth of high yield V-SWNTs. This is not understood currently and fine tuning of $O_2$ concentration might be needed to enhance SWNT growth in thermal CVD processes.

We also note an effect of H-species on the diameter distribution of SWNTs synthesized by CVD. The blocking of SWNT formation by reactive H-species appears to be more pronounced for smaller diameter tubes. Theoretically, smaller SWNTs are more susceptible to hydrogenation by H-species due to higher tube curvature and a higher $sp^3$ formation tendency.(14) In our H-plasma etching experiments, we observed the trend that smaller SWNTs tended to be attacked preferentially over larger ones, as seen in the AFM images in Fig. 4. This is also consistent with the observation that oxygen-free hydrocarbon CVD(17, 18) generally produce large (2-3nm) SWNTs with few tubes $\leq$ 1nm (when particles of various sizes <~4-5nm exist). In stark contrast, without hydrogen, CO based CVD methods are well known to produce abundant SWNTs in the 0.7-1.5nm range. Our $O_2$ assisted $CH_4$ PECVD presented here also synthesizes abundant SWNTs in the 1nm range (see Raman data in Fig. 1e). The growth of vertical MWNTs should be less affected by H-blocking due to the higher stability of larger tubes(14) and have been readily achieved by hydrocarbon CVD without any oxygen assistance.(2-6)

Lastly, we present a novel method for obtaining vertically aligned SWNTs on a wide range of substrates including metals and polymers. This goal has been elusive thus far due to the incompatibility of many types of substrates with the high growth temperature of SWNTs, but could be the key to the utility of aligned SWNT materials. Our method is simple and involves 'lifting off' the V-SWNT by using HF (1% for 10s) to etch the underlying $SiO_2$ layer and subsequently free-floating the V-SWNT film on a water surface (Fig. 6a). Following lift-off, the V-SWNTs can be transferred to other substrates coated with an interfacial thin polymer (50nm polymethylmathacrylate PMMA) layer for adhesion (Fig. 6b). After transfer, the substrate is heated to >190°C, well above the glass transition temperature of PMMA (~105 °C) for melting of the polymer layer and 'gluing' the substrate to the ends of the SWNTs in the vertical film. This affords strongly adhering vertical SWNTs on various substrates including Cu (Fig. 6b inset), polymers and glasses. The V-SWNT films thus derived are robust and do not lift off from substrates even after immersion in ethanol or acetone solvents. This development may greatly expand the utility of V-SWNTs. For instance, it has been suggested that vertically aligned SWNT films can be used as a thermal interface material for heat conduction and dissipation of microelectronics chips. A low temperature process is needed to form V-SWNTs on these chips without harming the pre-formed circuitry. Our room temperature SWNT transfer approach meets this challenge.

**Conclusions.**

We have presented a molecular oxygen assisted plasma-enhanced CVD growth of high yield of vertically aligned SWNTs at the full 4-inch wafer scale. Various control experiments and knowledge from previous diamond PECVD work revealed the negative effect of hydrogen species to the formation and growth of SWNTs as well as etching effects of hydrogen plasma to pre-formed SWNTs. The key role played by oxygen in our high yield SWNT growth is to balance C and H radicals, and specifically to provide a C-rich and H-deficient condition to favor the formation of $sp^2$-like graphitic structures. This understanding has important implications to SWNT growth in general by various methods. With the addition of suitable amount of oxygen to suppress H species, we expect that various types of PECVD setups can produce SWNTs at ultra-high yield and efficiency.  Further, we present a method to form V-SWNT films for the first time on any desirable substrate (including metals and plastics) with strong interfacial adhesion.

**Acknowledgements**.  This work was partly supported by Stanford GCEP.

**Figure Captions**

**Figure 1.** Molecular $O_2$ assisted synthesis of high yield vertical single-walled carbon nanotubes. (a) AFM image of a sub-monolayer of Fe nanoparticles (1-2nm in topographic height) formed on a $SiO_2$/Si wafer used for the synthesis. (b) Optical image of a visually black vertical SWNT film grown on a full 4-inch wafer. The tube-free region at the lower right of the wafer was due to clamping during deposition of Fe film. (c) A SEM image showing the slanted view of a V-SWNT film grown on $SiO_2$/Si. (d) A SEM side-view of a V-SWNT film. (e) Raman spectrum of a SWNT film. Laser excitation wavelength is 785 nm. (f) TEM image of SWNTs after being sonicated off of the $SiO_2$ and then dispersed onto a TEM grid. Note that using excessive $O_2$ (>~2%, below the ignition point of 4%) give zero yield of SWNTs due to oxidation of nanotubes and should be avoided.

**Figure 2.** Molecular $O_2$ assisted growth of vertical SWNT towers and sheets with down to sub-micron dimensions. (a) A SEM image showing SWNT towers with various widths (20 μm, 5μm, 1μm, 500 nm, 300 nm from left to right of the front region of the image) and vertical SWNT sheets (20 μm, 5 μm, 1 μm, 500 nm, 300 nm, 100 nm thick from top to bottom of the upper part of the image) after 30 min growth. (b) An AFM image of the patterned catalyst strips (bright 300nm and 100nm wide regions respectively) comprised of densely packed Fe nanoparticles used for the growth of the 300 nm and 100 nm thick vertical SWNT sheets (pointed by arrows) in (a). (c) An AFM image of two of the patterned catalyst squares (300 nm in width) used for the growth of the smallest towers (pointed by an arrow, tilted due to high aspect ratio) in (a). (d)&(e) Two SEM images of square, circular towers and lines of V-SWNTs, each from different growths than the sample in (a), to show the reproducibility of the synthesis.

**Figure 3.** Roles of hydrogen and oxygen. (a) SEM of vertical SWNTs grown with $CH_4/O_2$ (Partial pressure of $O_2 = 0.8\%$). (b) Very low yield of SWNTs grown in $CH_4/H_2$ (Partial pressure of $H_2 = 7.4\%$). These control experiments were performed on the *same* catalyst/substrate and total pressure. (c) A schematic illustration of the proposed role of oxygen species in hydrocarbon based synthesis of SWNTs. Scavenging of reactive H-species by oxygen shuts off (shown by the cross) the negative H-effect to SWNT growth. We suggest that this desirable oxygen effect happens at the catalyst particle site where oxygen blocks H-radicals from reacting with $sp^2 C$ networks formed on the metal seed particle. Inset: optical emission spectrum (OES) recorded under our optimum V-SWNT growth condition with $CH_4/H_2/O_2$ showing clear OH emission at 308.9 nm. The measurement was carried out using a S2000 miniature fiber optic spectrometer (Ocean Optics Inc.). UV grade quartz fiber was used to guide the plasma emission to the spectrometer.

**Figure 4.** Hydrogen attacking of SWNTs. AFM images of nanotubes on a substrate recorded (A) before and (B) after $H_2$ plasma treatment (5% in Argon, total pressure 0.5 torr, RF power 20W) at 500°C for 10 mins. The treatment was carried out in the same chamber used for growth of the V-SWNTs with only the $H_2/Ar$ gas flow. The after etching image in (B) clearly shows that some of the nanotubes in (A) were etched by H-plasma. SWNTs are also found to be etched by H-plasma at room temperature (data not shown)

**Figure 5.** Effect of hydrogen in regular thermal CVD growth. SEM images (scale bar 500nm) of nanotubes grown from silica-supported Fe/Co/Mo catalysts deposited on substrates by regular thermal CVD using ethanol as the carbon feedstock in the presence of a $H_2$ concentration of (A) 1.9 % and (B) 9.6 % respectively. 300 sccm of forming gas (3% $H_2$ in Argon) was bubbled through EtOH held at -9C in both growths. In addition to the bubbling forming gas, (A) had an additional 170 sccm of dry Argon flow, and (B) had 35 sccm dry Argon and 135 sccm dry $H_2$

flows. The particles seen in the images are silica with supported catalytic metal species. The catalyst is made of Fe:Co:Mo (molar ratio 1:1:0.2) acetate salts dissolved in anhydrous ethanol and sonicated with silica (0.72mmol metal/1 g Silica). The silica catalyst was deposited on a large silicon piece via spin coating. After spin-coating, the chip was cut into two pieces, and one was used for (A) and the other for (B). The yield of nanotubes is significantly higher in (A) than in (B); Surface 'sheet' resistance for (A) and (B) is 25K$\Omega$ and 1.2M$\Omega$ respectively.

**Figure 6.** How to form vertical SWNT films on a wide range of substrates. (a) Photograph of a V-SWNT film free-floating on water after lifted-off from a $SiO_2$/Si substrate by HF etching of the $SiO_2$ layer underlying the SWNTs. Right panel: a schematic drawing of the free-floating SWNT film with nanotubes held together by van der Waals interactions. (b) SEM image of a vertical SWNT film after transferring onto a copper substrate with a thin polymer binding layer at the Cu-SWNT interface. Inset: a photograph showing a vertical SWNT film (black) on Cu. Right panel: a schematic drawing depicting the vertical nanotube film and the Cu interface.

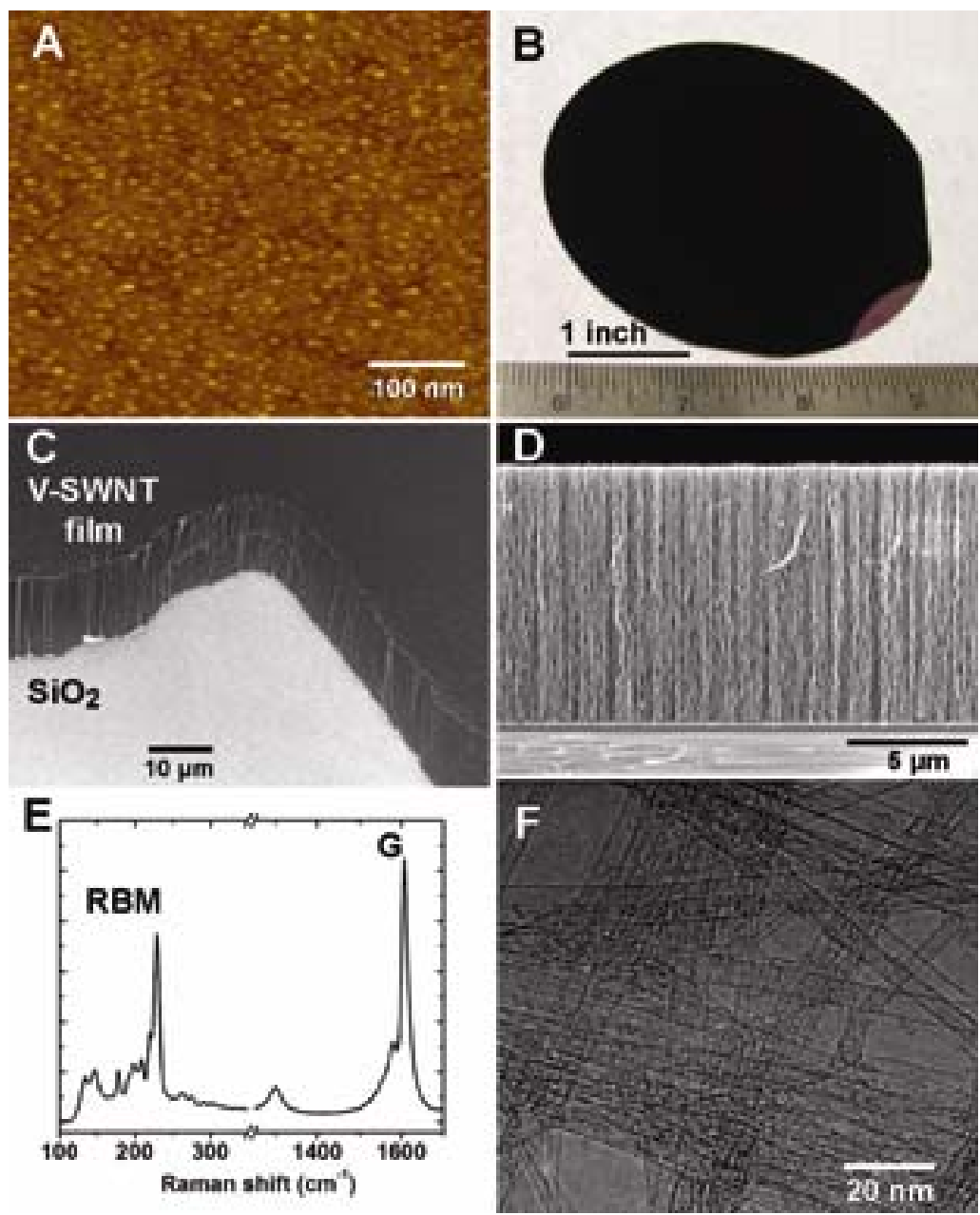

Figure 1

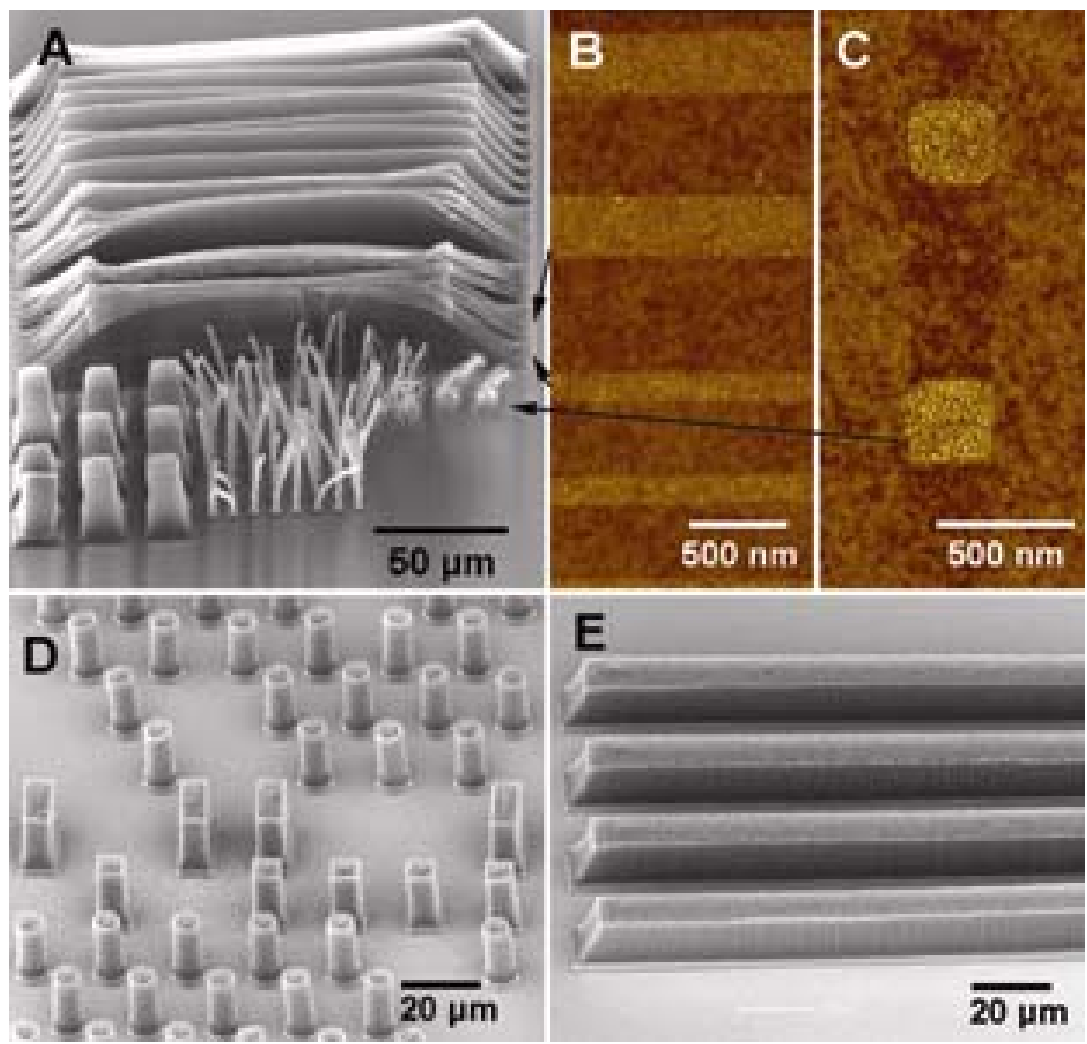

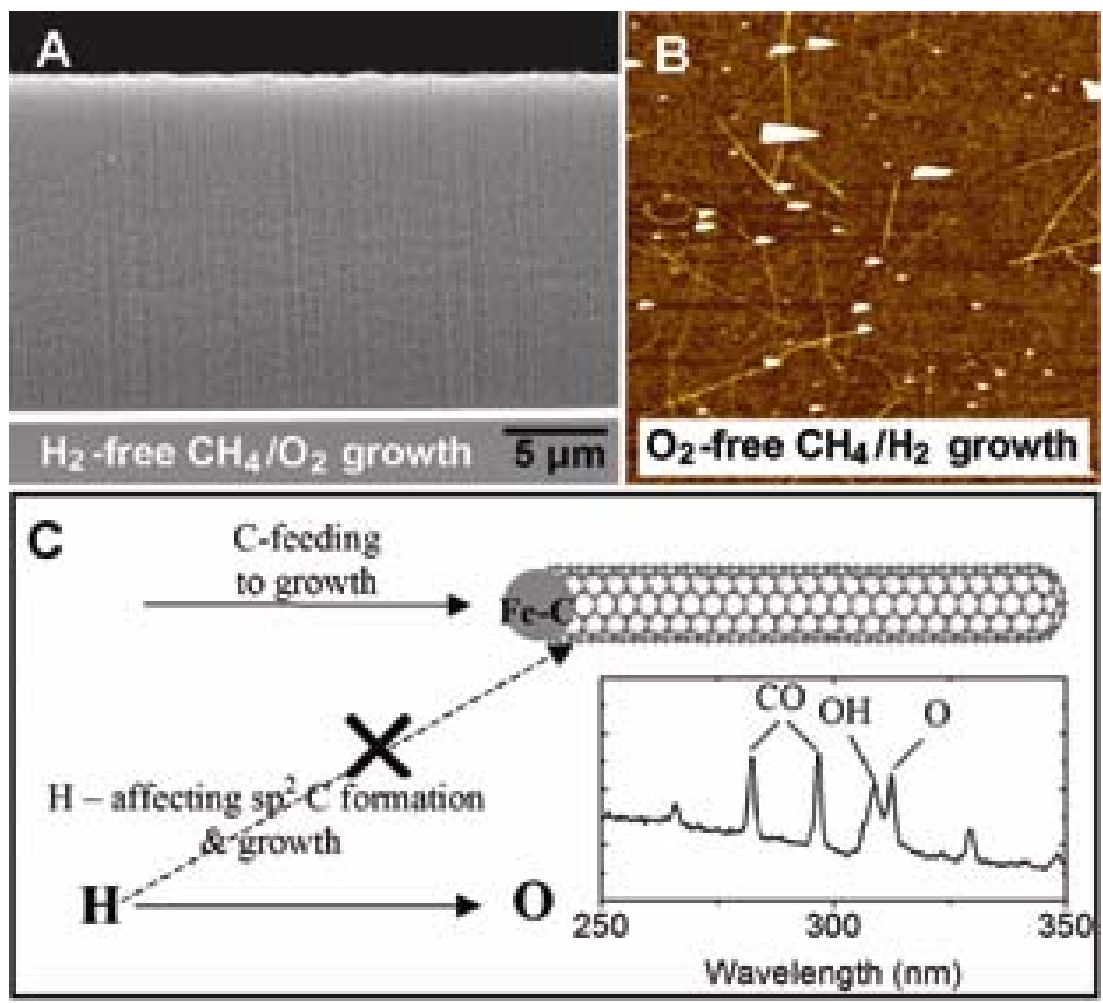

Figure 3

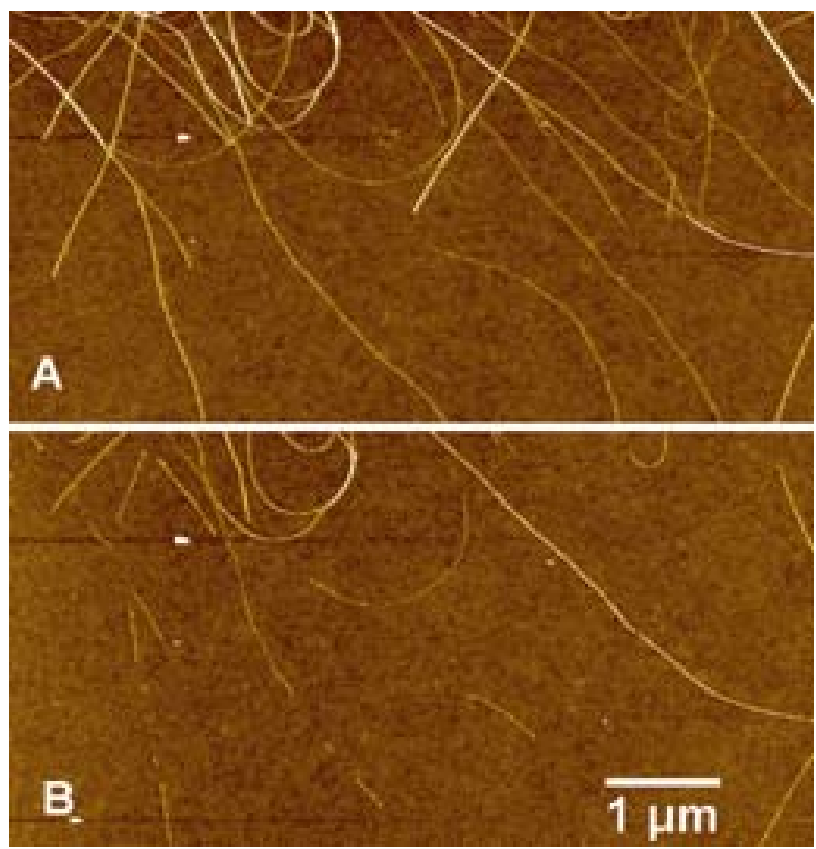

A

B

1 µm

Figure 4

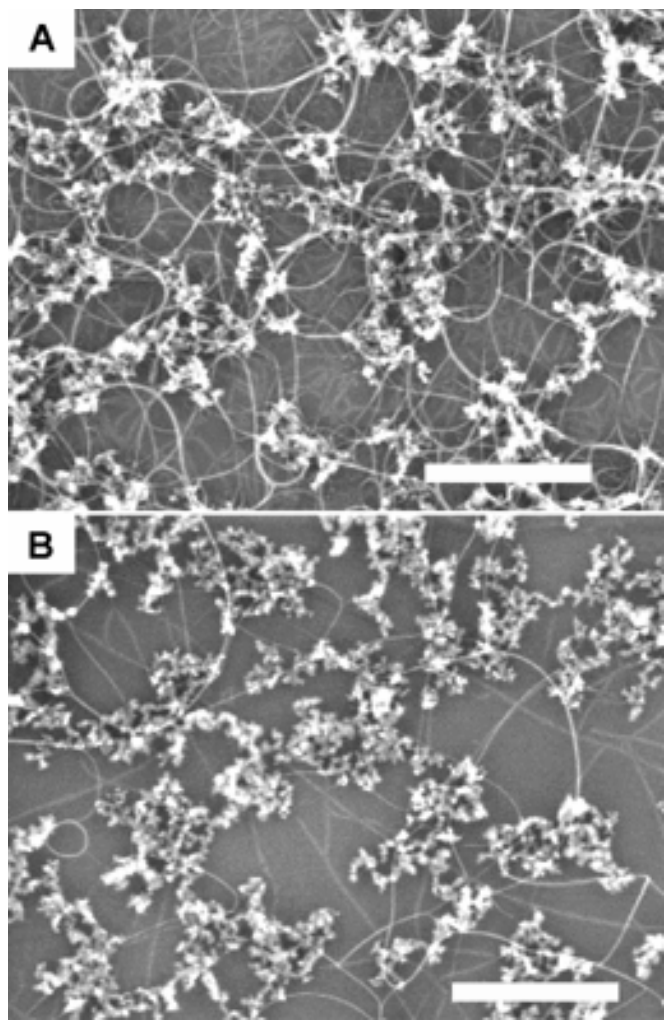

Figure 5

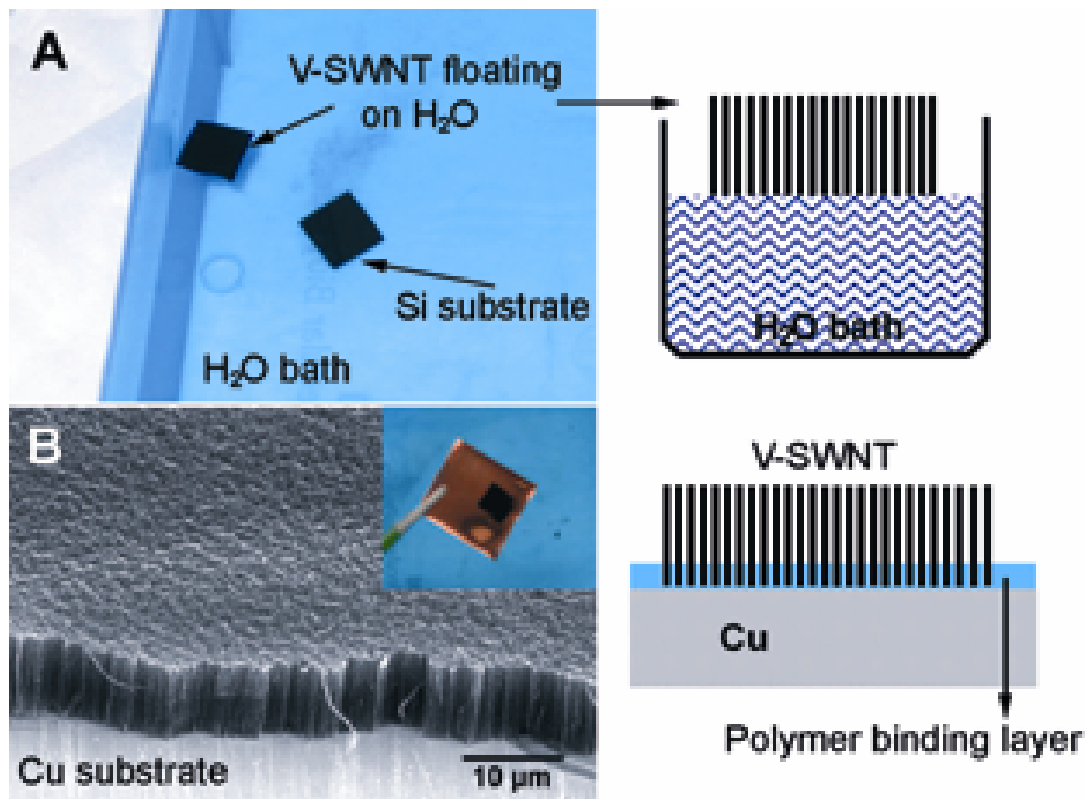

## Supplementary Information

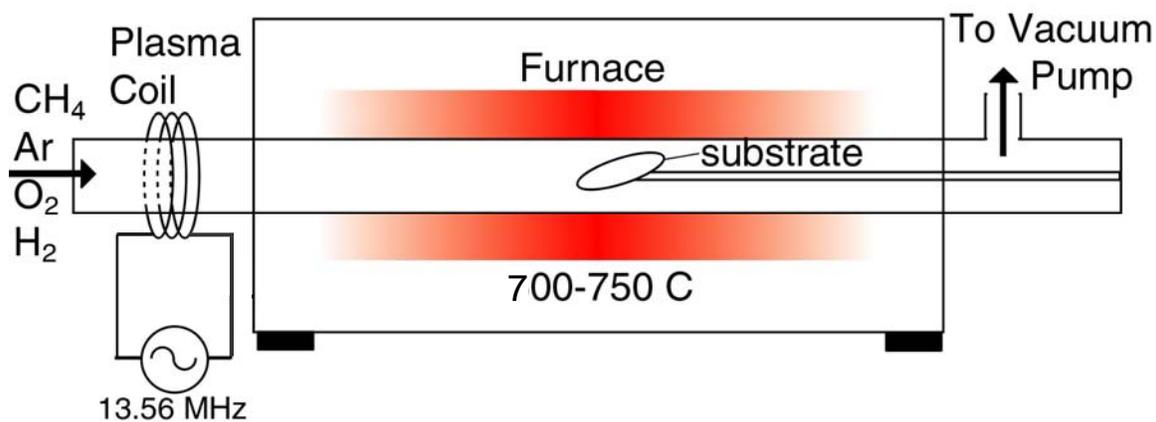

**Supplementary Figure 1: The 4-inch CVD system used for the vertical SWNT synthesis.**  A schematic drawing of the system used for our nanotube synthesis.

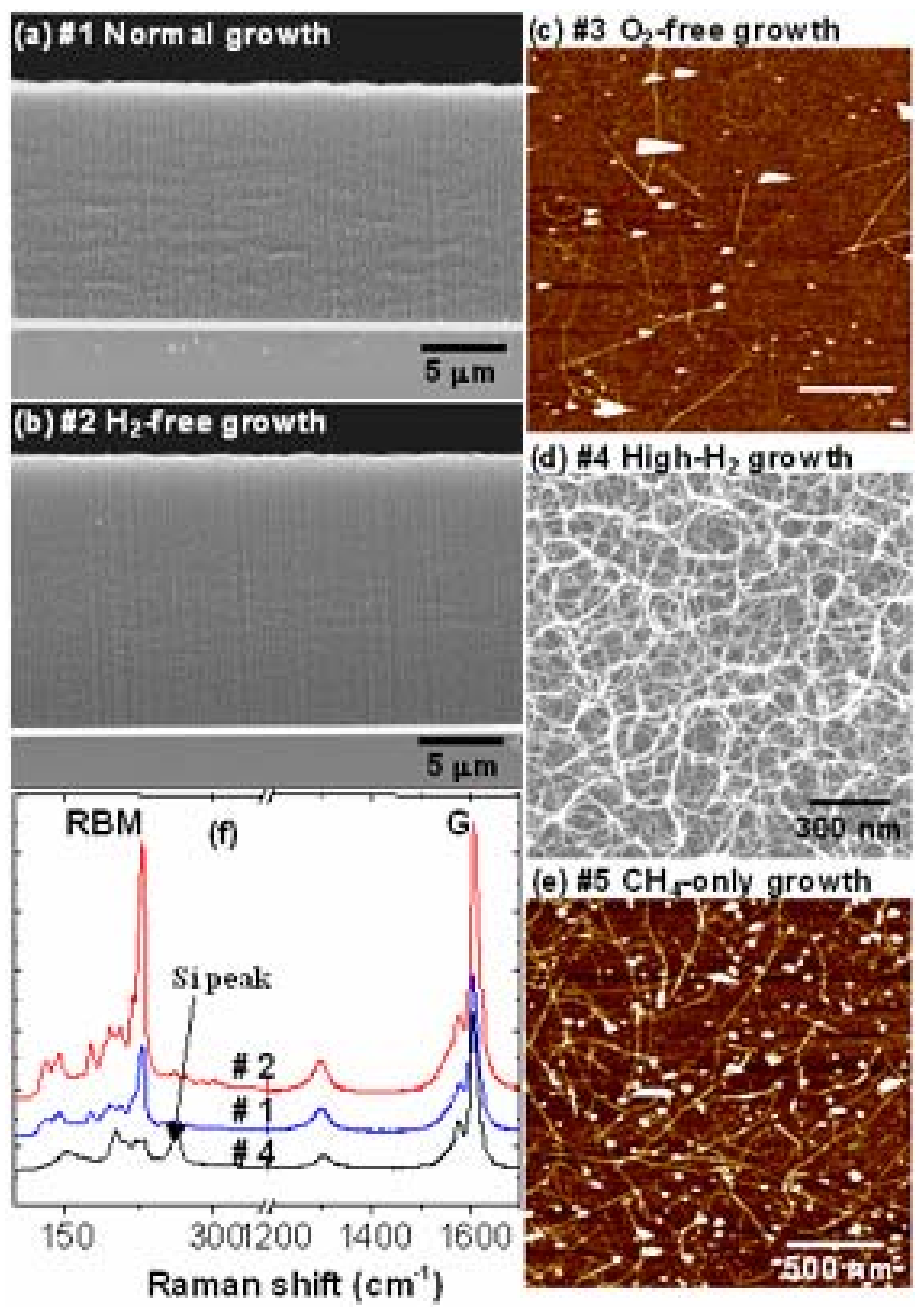

**Supplementary Figure 2: Comparison of growth results with various O$_2$ and H$_2$ conditions.** (a)-(e) Growth results (SEM or AFM images) under various conditions. (f) Raman spectra obtained with the various samples except for #3 and #5 due to the low SWNT yield. All of the growth experiments were carried out on identically prepared substrates with the same sub-monolayer Fe nanoparticle catalyst.  All growth runs were carried out at the same temperature (720C) and total pressure (0.34 torr) and plasma power (70W) for the same amount of time of 10 min.  Specific growth conditions are as follows.

Growth Run #1: Regular (normal) growth condition result with $CH_4/H_2/O_2$. Partial pressure $H_2 : O_2 = 12\% : 1\%$ (very high yield, SWNTs grown into vertical films).

Growth Run #2: $H_2$-free growth condition (no $H_2$ flow used in growth). $CH_4/O_2$ only. Partial pressure of $O_2 = 0.8\%$ (very high yield, SWNTs grown into vertical films). Removal of $H_2$ from the regular growth condition can still afford high yield growth, after adjusting down $O_2$ partial pressure. $H_2$ is not essential in our regular condition but can be added for balancing purposes.

Growth Run #3: $O_2$-free growth condition (no $O_2$ flow used in growth.) $CH_4/H_2$ only. Partial pressure of $H_2 = 7.4\%$ (very low yield of SWNT with $H_2$ but no $O_2$ in growth).

Growth Run #4: A $CH_4/H_2/O_2$ growth condition with high $H_2$ concentration. $H_2 : O_2 = 22\% : 1\%$ (> $12\% : 1\%$ in regular optimum growth condition in #1). (Adding excessive $H_2$ gives lower yield than #1and fails to produce vertically packed SWNTs. Rather, a mat of SWNTs lying on the substrates was formed).

Growth Run #5: $CH_4$ only growth (yield higher than $CH_4/H_2$ growth in #3, but cannot grow packed V-SWNTs).

General Trends: (1) Vertical SWNTs were never grown without $O_2$ for the $CH_4$ PECVD method (in e.g., run #3,#4,#5). (2) Higher $H_2$ conditions give lower yield of SWNTs. For $CH_4/H_2/O_2$ conditions, increasing $H_2$ gives lower yield (e.g, run # 4 vs. #1). $CH_4/H_2$ (#3) yield lower than $CH_4$ alone (#5) growth.